\documentclass[a4paper,english]{aa}
\usepackage{times}
\usepackage[T1]{fontenc}
\usepackage{amsmath}
\usepackage{graphicx}
\usepackage{amssymb}
\usepackage{natbib}
\usepackage{babel}
\usepackage{xspace}
\bibpunct{(}{)}{;}{a}{}{,}

\newcommand{\usnob}   {\mbox{USNO-B1.0}\xspace}

\newcommand{\dias}{\mbox{DAML02}\xspace}

\newcommand{\mc}[3]{\multicolumn{#1}{#2}{#3}}
\defcitealias{khea12}{paper~I}
\begin{document}

\title{Global survey of star clusters in the Milky Way}

\subtitle{II. The catalogue of basic parameters}

\author{N.V.~Kharchenko \inst{1,2,3} \and
        A.E.~Piskunov \inst{1,2,4} \and
        E.~Schilbach \inst{1} \and
        S.~R\"{o}ser \inst{1} \and
        R.-D.~Scholz \inst{2} }

\offprints{R.-D.~Scholz}

\institute{
Astronomisches Rechen-Institut, Zentrum f\"ur Astronomie der Universit\"at
Heidelberg, M\"{o}nchhofstra\ss{}e 12-14, D--69120 Heidelberg, Germany
\and
Leibniz-Institut f\"ur Astrophysik Potsdam (AIP), An der Sternwarte 16, D--14482
Potsdam, Germany\\
email: rdscholz@aip.de
\and
Main Astronomical Observatory, 27 Academica Zabolotnogo Str., 03680 Kiev,
Ukraine
\and
Institute of Astronomy of the Russian Acad. Sci., 48 Pyatnitskaya Str., 109017
Moscow, Russia
}

\date{Received 17 July 2013 / Accepted 9 August 2013}

\abstract
{Although they are the main constituents of the Galactic disk population, for half of the open clusters in the Milky Way reported in the literature nothing is known except the raw position and an approximate size.}
{The main goal of this study is to determine a full set of uniform spatial, structural, kinematic, and astrophysical parameters for as many known open clusters as possible.}
{On the basis of stellar data from PPMXL and 2MASS, we used a dedicated data-processing pipeline to determine kinematic and photometric membership probabilities for stars in a cluster region.}
{For an input list of 3784 targets from the literature, we confirm that 3006 are real objects, the vast majority of them are open clusters, but associations and globular clusters are also present. For each confirmed object we determined the exact position of the cluster centre, the apparent size, proper motion, distance, colour excess, and age. For about 1500 clusters, these basic astrophysical parameters have been determined for the first time. For the bulk of the clusters we also derived the tidal radius. We estimated additionally average radial velocities for more than 30\% of the confirmed clusters. The present sample (called MWSC) reaches both the central parts of the Milky Way and its outer regions. It is almost complete up to 1.8 kpc from the Sun and also covers neighbouring spiral arms. However, for a small subset of the oldest open clusters ($\log t \gtrsim 9$) we found some evidence of incompleteness within about 1~kpc from the Sun.}
{}


\maketitle

\section{Introduction}\label{sec:intro}

Star clusters are the main building blocks of the stellar populations in our Galaxy. They are are found in high numbers among different Galactic populations, and their astrophysical parameters can be determined with relatively high precision. In the literature about 4000 objects are currently known that can be regarded as Galactic star clusters of various types or candidates. However, about half of them are cited only with a name, approximate coordinates (R.A. and Decl.), and a rough angular size. We call all clusters mentioned in the literature, or more precisely, the clusters of our input sample (defined below)  \textit{named} clusters henceforth to distinguish this sample from our final sample of clusters for which we determined astrophysical parameters within \textit{our Milky Way Star Clusters (MWSC)} project.

We started \textit{the MWSC} survey a few years ago and aimed to build a comprehensive sample of Galactic star clusters with well-determined parameters, spatially complete enough to enable an unbiased study of the content and evolution of the star clusters of our Galaxy.  For a reliable cluster membership construction and  a reliable cluster parameter determination we used a combination of uniform kinematic and accurate near-infrared (NIR) photometric data gathered in the all-sky catalogue PPMXL \citep{ppmxl}. In the first paper \citep{khea12}, hereafter called \citetalias{khea12}, we introduced the survey, explained the underlying motivation, provided a short review of similar studies, described the observational basis of the survey and our data processing pipeline, and presented preliminary results obtained in the second Galactic quadrant. The present paper summarises the results of the full survey carried out for a compiled list of all named clusters, covering the whole sky. The MWSC catalogue of 3006 clusters, the corresponding stellar data with membership probabilities, as well as supplementary material on the full input list of clusters will be available from the CDS\footnote{ftp://cdsarc.u-strasbg.fr}. We do not plan, however, to restrict the survey to named clusters only. Our second aim is extending the cluster sample by detecting hitherto unknown clusters on the basis of the PPMXL catalogue. This work is still in progress, and we will present the results in a following paper.

In Sect.~\ref{sec:stat} we briefly describe the basic input data set and the pipeline of the cluster parameter determination and give the current general statistics of the outcome. In Sect.~\ref{sec:param} we characterise the derived cluster parameters. In Sect.~\ref{sec:smpl} we discuss the general properties of the MWSC sample of star clusters. Sect.~\ref{sec:conc} summarises the results achieved so far.

\section{Survey constituents and output statistics}\label{sec:stat}

\subsection{Observational basis}\label{sec:input}

The basic stellar data for our study were taken from the all-sky catalogues PPMXL \citep[]{ppmxl} and 2MASS \citep{cat2MASS}. PPMXL gives coordinates $\alpha,\,\delta$, and proper motion components  $\mu_\alpha$ and $\mu_\delta$ in the ICRS, and low-accuracy photometry from  \usnob \citep[]{usnob1} for about 900~million objects down to $V\approx20$. For some 400~million
entries the catalogue  contains accurate $J,H,K_s$ magnitudes from 2MASS \citep{cat2MASS}. For the current survey we used an improved merged version of these catalogues called hereafter 2MAst, see \citetalias{khea12} for a  description of its construction. We used 2MAst to confirm clusters from our input list and to determine cluster parameters in the astrometric and photometric systems that are homogeneous over the whole sky.

\begin{table}
  \caption{Comparison of the COCD and MWSC surveys.}
  \label{tbl:c&m}
  \begin{tabular}{lll}\hline
                          & COCD            & MWSC            \\
\hline
  Basic catalogue         & Hipparcos+Tycho  & PPMXL            \\
  The tool catalogue      & ASCC-2.5        & 2MAst            \\
  Number of stars         & 2.5 mln.        & 470 mln.         \\
  Limiting magnitude      & $V=12.5$        & $K_s=15.3$       \\
  Basic stellar data      & $\alpha,\delta,\mu,BV$  & $\alpha,\delta,\mu,JHK_s$\\
  Additional stellar data & $JHK_s,Sp,RV$   & $BV,Sp,RV$       \\
  Limiting distance       & ~8 kpc          & $>20$ kpc          \\
  Completeness            & 0.85 kpc        & 1.8 kpc            \\
  \hline
  \end{tabular}
\end{table}

In Table~\ref{tbl:c&m} we compare the basics of the current survey with our previous work on open clusters \citep{clucat,newc109}, here referred to as COCD (catalogue of open cluster data), which was based on the shallower ASCC-2.5 catalogue. The comparative depth of the new survey has increased by at least 3 mag by considering the transitions from the optical to the infra-red. This provides the respective increase of the limiting and completeness distances. Including the the NIR photometric data has, as a consequence, led to an increased relative weight of the clusters of extreme age (both young, normally heavily reddened, and old, which are relatively faint in the optical).

The target list was compiled from sources available in the literature. As the primary source we used the data from COCD. For additional optical clusters and associations the data were taken from the catalogue of \citet{daml02}, called hereafter as \dias (version 3.1, 24/nov/2010), and from \citet{meldam09}. For clusters detected in the NIR the information came from \citet{bicaea03}, \citet{bicadb03}, \citet{dubisb03}, \citet{froeb07}, \citet{froea10}, \citet{bukea11}, and other sources. We described the input list in detail in \citetalias{khea12}. Furthermore, we included embedded and globular clusters taken from the catalogues by \citet{lala03} and \citet{harris96}[edition 2010]\footnote{http://www.physics.mcmaster.ca/resources/globular.html} in our target list.

\subsection{Pipeline}\label{sec:pipe}

The pipeline has been described in detail in \citetalias{khea12} and here we only refer to its basic features. The main purpose of the pipeline is i) to reveal clusters from the fore-/background using kinematic, photometric, and spatial criteria, ii) to construct a list of cluster members, and iii) to determine basic cluster parameters. For each cluster, the direct outcome of the pipeline were the improved coordinates of the cluster centre and the apparent sizes of different  morphological parts of the cluster, the average proper motion, the distance, the reddening, and the age. As a theoretical basis, we used recent Padova stellar models with isochrones computed with the CMD2.2 on-line server\footnote{http://stev.oapd.inaf.it/cgi-bin/cmd}, whereas the pre-MS isochrones were based on the models of \citet{siessea00} transformed by us to the $JHK_s$ photometric system. In the pipeline, we used several diagrams of kinematic (vector point diagram, magnitude-proper motion relation) and photometric (colour-magnitude, two-colour and $Q_{JHK_s}$-colour diagrams) data. The membership probabilities of stars in the diagrams were determined from their location with respect to the reference sequences (represented by isochrones in photometric diagrams, or the average cluster proper motion in kinematic diagrams), which themselves depend on the cluster parameters we intended to find. Hence, this required an iterative approach, allowing us to successively improve both cluster membership and cluster parameters. As an initial approximation we either used data from the input list or made our own estimates by eye if the input parameters were obviously incorrect or not available. As a rule, the process converged after a few iterations. The pipeline provided spatial, kinematic, and photometric membership probabilities for each star in the sky area around a cluster. In total, about 64 million stars were retrieved. For each confirmed cluster (see below), a coherent set of basic cluster parameters was determined.

\begin{table}
\caption{Statistics of MWSC objects}
\label{tbl:objstat}
\begin{tabular}{lrrl}
\hline
\noalign{\smallskip}
Parameter&Number& \% &Note \\
\hline
\noalign{\smallskip}
Surveyed objects            &3784 & 100 &  \\
Confirmed objects           &3006 &  79&  \\
Unconfirmed objects       & 778 &  21&  \\
~~~~dubious objects         & 399 &  11& 1\\
~~~~not established objects & 158 &   4& 2\\
~~~~duplicated objects      & 221 &   6& 3\\
\hline
\noalign{\smallskip}
\end{tabular}\\
\begin{scriptsize}
1:~neither membership nor cluster parameters can be determined;
2:~not seen in 2MAst;
3:~already identified under a different name, or is part of another cluster.\\
\end{scriptsize}
\end{table}

\begin{table}[b]
\caption{Cluster types in the MWCS survey}
\label{tbl:cls}
\begin{tabular}{lrrr}
\hline
\noalign{\smallskip}
Class&Number&Known&Candidate \\
\hline
\noalign{\smallskip}
Globular clusters    & 147  & 142   &   5        \\
Associations         &  51  &  21   &  30        \\
Open clusters:       & 2808 & 2808  &   -        \\
~~~~remnants         & 389  &  221  & 168        \\
~~~~with nebulosity  & 132  &  132  &   -        \\
~~~~moving groups    &  19  &   19  &   -        \\
\hline
\noalign{\smallskip}
\end{tabular}
\end{table}                    

The final information on membership was then also used to determine of other characteristics of a cluster, such as parameters of the \citet{king62} density profile and the radial velocities (RVs). King parameters (core and tidal radii, and normalisation factors) were computed for almost all clusters. The RV of a cluster was determined by averaging  RVs of individual cluster members, for which these data were available in the literature. Altogether, RVs were obtained for about 30\% of the MWSC objects.

Since open clusters belong to a relatively young Milky Way population, their metallicities are high (of the order of the solar value), and, as the stellar model computations show (see for example the Padova isochrones of the same age and different $Z$), their variations only mildly influence the derived cluster parameters such as age and distance. Therefore we adopted the isochrones with solar metallicity ($Z=0.019$) to determine the parameters of open clusters. When a metallicity measurement was found  in the literature for a given cluster, its value was copied to our catalogue. Since we used the data from the bibliographic data collection of \dias as the main source of open cluster metallicities, the MWSC metallicities, unlike other cluster parameters, should be regarded as highly heterogeneous. For globular clusters, the  distances, reddenings and ages determined from colour-magnitude diagrams (CMDs) strongly depend on the adopted metallicity, which means that  the approach with solar metallicity is not applicable. Moreover, for the majority of globular clusters we observed only the brightest red giant members in 2MAst. The steepness of the giant branch makes it difficult to estimate their distances reliably. Therefore, we adopted the published distances and reddenings from the \citet{harcat10} catalogue, and determined ages and metallicities from the theoretical isochrones that provided the best fit to the observed CMD.

\begin{table}
\caption{Statistics of cluster parameters in the MWSC, the first seven parameters
are called basic parameters, the other three additional parameters}
\label{tbl:pstat}
\begin{tabular}{lrrr}
\hline
\noalign{\smallskip}
Parameter&\mc{3}{c}{Number} \\
\hline
\noalign{\smallskip}
                 & MWSC &  1st time & \% \\
\hline
Coordinates      & 3006 &     0    &  0 \\
Membership       & 3006 &  1217    & 40 \\
Apparent radius  & 3006 &    20    & <1 \\
Proper motions   & 3006 &  2110    & 70 \\
Distance         & 3006 &  1386    & 46 \\
Reddening        & 3006 &  1555    & 52 \\
Age              & 3006 &  1594    & 53 \\
                 &      &          &    \\
Tidal parameters & 2961 &  1276    & 43 \\
Radial velocity  &  953 &   243    & 25 \\
Metallicity      &  386 &     0    &  0 \\
\hline
\noalign{\smallskip}
\end{tabular}\\
\end{table}

\subsection{Object statistics}\label{sec:res}

In total the input list of named clusters comprised 3784 entries. In our analysis, however, we did not find real counterparts for 21\% of the targets in 2MAst.  The corresponding statistics is given in Table~\ref{tbl:objstat}. About one half of the uncomfirmed targets seem to be dubious objects that show no reasonable sequences either in the kinematic or photometric diagrams, therefore we concluded that they are not physical clusters. Therefore, no membership probability was provided for the stars in the corresponding sky areas. Another 4\% of entries from the input list were not seen in 2MAst since they are too faint for this catalogue. Moreover, very many targets in the input list turned out to be double or even multiple entries. In different sources in the literature, the same object was referred to by different names and/or with slighly different positions. We kept unconfirmed entries in the output list and marked them with corresponding flags and notes.

About 79\% of the objects considered were confirmed as real clusters. In total, the 3006 confirmed clusters contain about 400000 stars that are most probably cluster members, that is, stars with kinematic and photometric membership probabilities higher than 60\%. On average, this corresponds to about 130 stars per cluster (to be compared with the about 20 highly probable members per cluster in COCD).

The MWSC survey includes clusters of different types, which are listed in Table~\ref{tbl:cls}. There are 142 globular clusters from \citet{harcat10}, which make up about 90\% of entries of this catalogue. Five new globular cluster candidates are star clusters observed in the direction of the Galactic centre. We classified them as globular clusters because of their CMDs. We will discuss globular clusters from the MWSC survey in more detail in a dedicated paper (Kharchenko et al., 2013, in preparation).

Associations in the MWSC are defined as plain assemblies of early-type stars that appear like star clusters, but do not show a pronounced central concentration in the density profile of the member distribution on the sky. We stress here that both our pipeline and the basic input data are highly unsuitable for a study of extended star-forming regions with multi-centre structures, which are usually referred to by the term association. Therefore, many classical associations cannot be processed with our pipeline, and were flagged as dubious objects in the MWSC. For 21 compact cluster-like assosiations from \citet{meldam09} the pipeline provided acceptable results however. We refer to targets as candidates in Table~\ref{tbl:cls} if they were classified as clusters in the original input list, but did not show a clear central concentration although they contained early-type stars. Finally, individual clusters observed within large and complex associations that successfully passed our pipeline were classified as confirmed open clusters.

Furthermore, we considered several sub-groups of open clusters (see Table~\ref{tbl:cls}). The descriptors ``cluster with nebulosity'', ``moving group'', and ``remnant'' were adopted from the original sources. We also marked old scarce clusters with irregular density profiles as  ``remnant candidates''.

\begin{table}
\caption{Comparison of the bacic cluster parameters in the MWSC survey and data in the literature}
\label{tbl:regrc}
\begin{tabular}{lcccr}
\hline
\noalign{\smallskip}
Parameter&$a$            &$b$&$\sigma$& $N_{obj}$ \\
\hline
\noalign{\smallskip}
 $\mu_x$, mas/yr &$\;\;\;0.22 \pm$ 0.06&  1.02 $\pm$ 0.01 & 1.23 &  882 \\
 $\mu_y$, mas/yr &$     -0.12 \pm$ 0.06&  1.00 $\pm$ 0.01 & 1.20 &  884 \\
 $(V-M_V)$, mag  &$     -0.33 \pm$ 0.08&  1.01 $\pm$ 0.01 & 0.46 & 1403 \\
 $\log d$, pc    &$     -0.11 \pm$ 0.02&  1.03 $\pm$ 0.01 & 0.07 & 1406 \\
 $E(B-V)$, mag   &$     -0.04 \pm$ 0.01&  1.01 $\pm$ 0.01 & 0.10 & 1410 \\
 $\log t$, yr    &$\;\;\;0.05 \pm$ 0.10&  0.99 $\pm$ 0.01 & 0.24 & 1262 \\
\hline
\noalign{\smallskip}
\end{tabular}
\end{table}                    

\section{Cluster parameter scope}\label{sec:param}

\begin{table}[b]
\caption{Estimated accuracy of the derived MWSC cluster parameters}
\label{tbl:pacc}
\begin{tabular}{lll}
\hline
\noalign{\smallskip}
Parameter&Internal error           &External error \\
\hline
\noalign{\smallskip}
 Proper motion         & $\pm0.5\, \mathrm{mas/yr}\,(\delta>-20^\circ)$ & $\pm0.8$~mas/yr\\
                       & $\pm0.8\, \mathrm{mas/yr}\,(\delta<-20^\circ)$ &              \\
 Distance              & $-$                                            & 11\%         \\
 $E(B-V)$              & $-$                                            & ~~7\%         \\
 Age$^*$               & 25\%, $(\log t<8.2)$                           & 39\%         \\
                       & 10\%, $(\log t>8.2)$                           &              \\
 Tidal radius          & 25\%                                           & $-$          \\
 Radial velocity$^\dag$& $\pm1.0$ km/s                                  & $-$          \\
\hline
\noalign{\smallskip}
\end{tabular}\\
\begin{scriptsize}
$^*$Only 637 clusters have derived internal errors in the catalogue.\\
$^\dag$75\% of clusters have $\varepsilon_{RV}< 5.0$ km/s (for more details, see text).\\ 
\end{scriptsize}
\end{table}                    

One of the main goals of our survey is to determine a uniform and homogeneous set of astrophysical cluster parameters for all named clusters.

\begin{figure*}[t]
\includegraphics[width=\hsize]{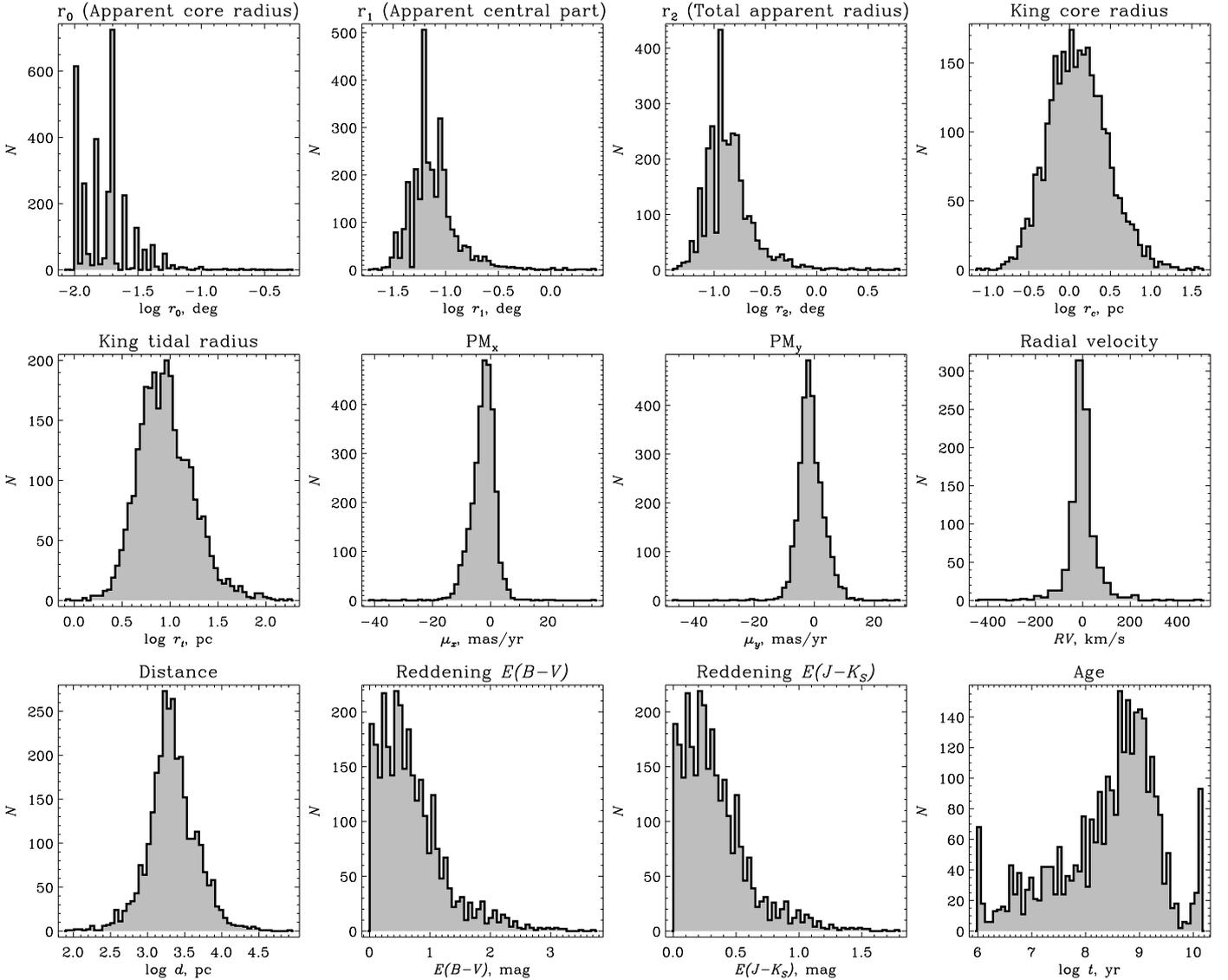}
\caption{Distribution functions of MWSC clusters with respect to the 
different astrophysical parameters determined in this paper.
}
\label{fig:dstr_all}
\end{figure*}

Our set of cluster parameters includes positions (coordinates of the centre, distance) that can be converted into 3D coordinates, structure data (apparent sizes of distinct cluster parts, tidal parameters), kinematic information (proper motions, RVs) related to 3D velocity components, and astrophysical parameters (age, reddening, and sometimes cluster metallicity). In Table~\ref{tbl:pstat} we show the number of clusters for which a given parameter was determined in the MWSC survey (column~2), and the number of clusters for which the parameter was estimated for the first time within the MWSC project  (column~3). Column~4 gives the corresponding  percentage of these clusters. Except for most of the metallicities, all parameters listed (including coordinates and apparent radii) were re-determined or newly determined in our study.

We distinguish in Table~\ref{tbl:pstat} basic parameters determined via the MWSC pipeline for all confirmed objects (first seven rows), and additional ones (last three rows) that required a different approach and were obtained for a fraction of clusters only. To determine the King tidal parameters we use information on the projected density distribution of cluster members, which is described in \citetalias{khea12} and the references therein. However, not all of the confirmed clusters follow King profiles. Especially for apparently large loose objects with poor radial density profiles one fails to fit a King model.
  
Originally, RVs were known for only 670 clusters of our sample  \citep{newrvcm,daml02}. To update the RV data, we used information on the membership and looked for the relevant RV measurements in CRVAD-2 \citep{newrvc2} and in SIMBAD. This led to 75 and 14 additional cluster RVs, respectively. For another 40 clusters, new RVs were determined by \citet{conrad13} from dedicated RAVE observations proposed earlier on the basis of COCD data. Additionally, we cross-matched about 63.6 million entries in 3006 MWSC cluster areas with the about 2.7 million spectroscopic measurements in the SDSS DR9 \citep{ahn12} and found only a small overlap of about 20000 measurements in 93 cluster areas. Only in 62 of these areas at least one cluster member had an RV measured in SDSS. Finally, we identified about one hundred cluster members as IRAS point sources with measured RVs \citep{bron96} and determined RVs for another 92 star clusters.

Metallicities for open clusters were taken from the \dias and \citet{conrad13}. For globular clusters we had to vary the cluster metallicity parameter to achieve the best agreement between the observed colour-magnitude diagrams and isochrones. The corresponding pipeline modifications for  globular clusters will be described in more detail in the forthcoming paper (Kharchenko et al., 2013, in preparation). From Table~\ref{tbl:pstat} one notes that the cluster metallicity is the most rarely known parameter in the MWSC survey.

In Table~\ref{tbl:regrc} we give the results of a comparison of the basic MWSC parameters with the corresponding data found in the literature. For the listed samples of $3\sigma$-clipped deviants, we applied the method of a least-squares bisector \citep{isobe90}, which provides the coefficients of a regression $y=a+b\,x$, and the standard deviation $\sigma$ around the bisector. The parameters $a$ and $b$ indicate the magnitude of a possible bias between the compared data sets, and $\sigma$ is the measure of the spread of the data around the regeression line. We found (Table~\ref{tbl:regrc}) that the scale coefficient $b$ is close to unity for all basic parameters, whereas the zero-point differences $a$ are formally significant.The bias in zero-points can be caused by uncertainities of our parameters as well as of the literature values. Taking into account the accuracy of the input data and the spread $\sigma$, the  differences are not dramatic however.

\begin{figure}[t]
\includegraphics[width=\hsize,viewport=0 0 360 360]{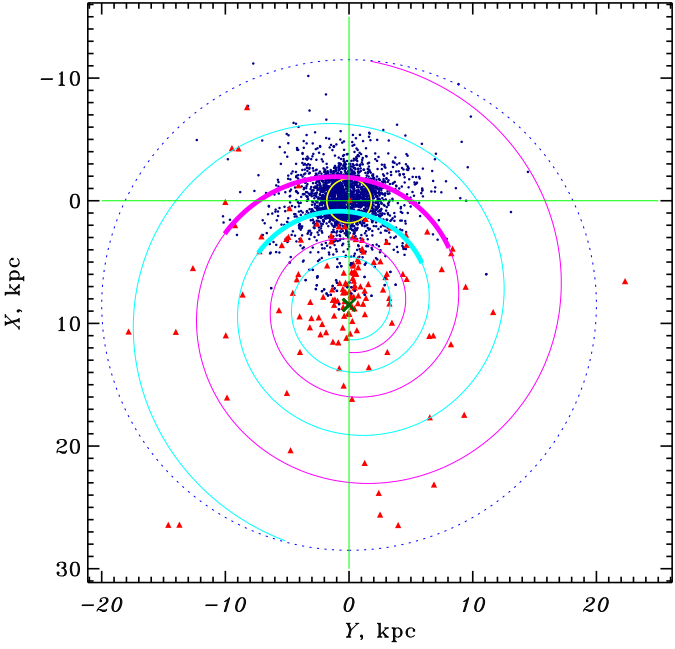}
\caption{Distribution of clusters of the MWSC survey
projected onto the Galactic $XY$-plane. Blue dots are open clusters and associations, red triangles mark globular clusters. The dotted circle shows the border of the Galactic disk (diameter $\sim$20~kpc) as indicated by distant open clusters. The thick solid cyan and magenta sections of the spirals indicate the position of local spiral arms as defined by the COCD clusters \citep{clupop}. They are formally extended to the edge of the disk with thin solid curves. The light (yellow) thick circle around the Sun with radius of 1.8 kpc marks the completeness limit of  the survey. A cross at $(X,Y)=(8.5,0)$ kpc indicates the position of the Galactic centre.
}
\label{fig:xy}
\end{figure}

Our estimates on the accuracy of the cluster parameters are given in Table~\ref{tbl:pacc}. There are two ways to estimate the achieved accuracy. For parameters that are average or fitted values (such as proper motions, RVs, or tidal radii, but also ages, if at least two stars were used in fitting to the evolved portion of the isochrone), one can estimate the spread of cluster members around the fit. This does not take into account the accuracy of the pipeline itself and gives us an internal estimate (lower limit) of the error only. For the overall characterisation of the internal errors of the full cluster sample, we used the most frequent value (the mode) of the error distribution. These values are shown in the second column of Table~\ref{tbl:pacc}. In the third column we show the external estimates of the derived parameter errors, which come from the comparison of our values with those published in the literature. For this comparison we used the dispersions given in the fourth column of Table~\ref{tbl:regrc}, and assumed that our data have at least the same accuracy as those from the literature. This comparison includes inaccuracies originating from our pipeline as well as from the technique/data used to determine the literature parameters. Therefore, these external errors represent an upper limit of our error estimates.

\begin{figure}[t]
\includegraphics[width=\hsize,viewport=0 0 525 430,clip]{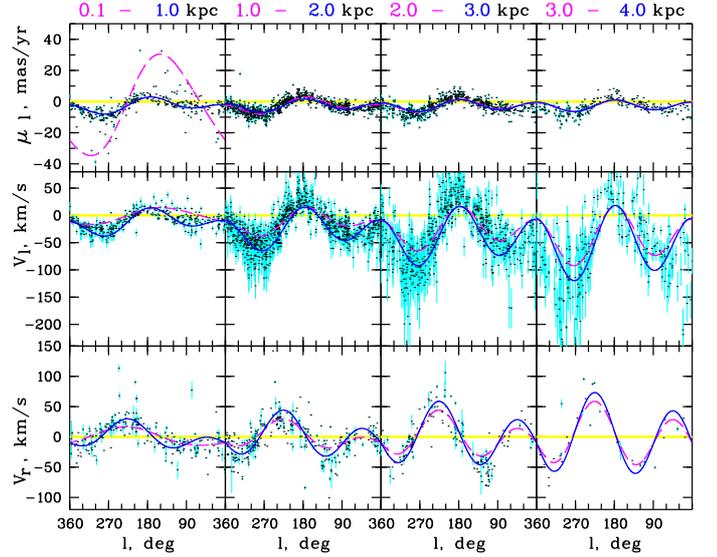}
\caption{Observed (derived) kinematic data of open clusters within $|b|\leq 20^\circ$ versus Galactic longitude for four ranges of projected distances $d_{XY}$ (as indicated at the top). The upper panels show proper motions, the middle panels tangential velocities, and the bottom panels RVs. Individual clusters are marked by thin horizontal lines with vertical error bars (in the $RV$ panel, error bars are given only for clusters whose $rms$-errors could be estimated). For a given distance range, the curves show the systematic velocity components due to solar motion and local Galactic rotation computed with COCD-based values \citep{clupop}. The dashed curves correspond to the contributions at the smallest distance of a given distance range, whereas solid curves show this contribution at the largest distance. For example, in the upper left panel the dashed curve shows systematic variations versus  longitude in proper motions expected for a cluster at 0.1~kpc from the Sun, whereas the solid curve shows these variations for a cluster at 1.0~kpc.
}
\label{fig:kinl}
\end{figure}

As one can see from Table~\ref{tbl:pacc}, the mean cluster proper motions
are
 accurate within 1 mas/y, and proper motions of northern clusters are
 somewhat more accurate than those of the southern clusters. Since distances
  and reddenings were determined from fitting the isochrones by eye
  to the observed colour-magnitude distributions, we were not able
  to determine the internal accuracy of the procedure. Therefore,
  we relied on the comparison with previously known values, which indicates
  a good quality of our derived distances and reddenings, even better
  than our preliminary quality assessment in
\citetalias{khea12}.
For clusters with well-observed turn-off points, we were able to estimate
 the spread of cluster members around the selected isochrone and
  consequently the internal uncertainty of the derived ages caused
    by this spread. We found that this uncertainty depends on age and
     is smaller for older clusters with more populated and clear
turn-offs.
      The external age errors are considerably larger than the internal
ones.
       One possible reason for this discrepancy are the different
theoretical
        isochrones applied. But a particularly strong impact on the age
         determination can be produced by wrongly including non-members
          (or vice versa by neglecting true members) above the turn-off.
           This occurs especially when only photometric
            membership is applied. Because we combined photometric and proper
     motion membership, our results are probably in general more
      reliable than those from pure photometric membership studies
       (as discussed in \citetalias{khea12}). Tidal radii
        are accurate to within 25\%. The distribution of the $RV$-errors
         peaks near 1~km/s and most of the clusters have
$\varepsilon_{RV}$ better
          than $\pm5$ km/s.
However, there is a caveat on the mean error estimation of the RV
of a cluster.
The uncertainty may be larger if radial velocities of only a few stars are
available,
and if these stars
also have low membership probability.
For 446 out of 953 clusters with RVs, more than one member
was measured, whereas the cluster RVs of 279 and 228 clusters were based on only 
one or an unknown number of members (if not given in the literature), 
respectively. The RVs of 183 clusters have no reported errors.

\begin{figure}[t]
\includegraphics[width=\hsize,clip=]{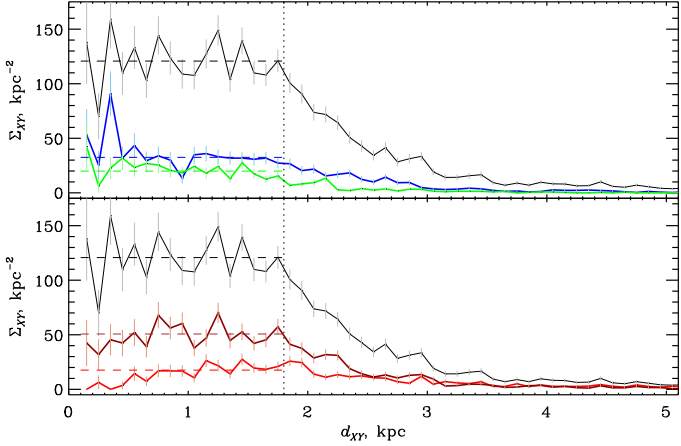}
\caption{Distribution of the surface density $\Sigma_{XY}$ of star clusters versus distance $d_{XY}$ from the Sun projected onto the Galactic plane. The density distribution of all clusters is given in black in both panels. In the upper and  lower panels we show the distributions of younger (blue: $\log t<7.9$, green: $\log t=7.9\dots8.3$) and older clusters (brown:
$\log t=8.3\dots9.0$, red: $\log t>9.0$), respectively. The dotted vertical line indicates the adopted completeness limit, the dashed horizontal lines correspond to the average surface density within the completeness limit of clusters in different age groups.}
\label{fig:sdprof}
\end{figure}

In Fig.~\ref{fig:dstr_all} we show histograms of cluster parameters determined in the MWSC survey. Note that these histograms characterise the distribution of all 3006 clusters in our catalogue, not the actual cluster population in the solar neighbourhood. We found that the typical angular radius $r_2$ of the clusters in our sample is $r_2\approx3\dots12$ arcmin, or that the typical tidal radii $r_t$ are between 5\dots10 pc. Most proper motions lie within $\pm 10$ mas/y, and the RVs within $\pm100$ km/s. Though the typical distance is about 2 kpc, a long tail is observed in the distribution of distant clusters up to more than 10 kpc from the Sun. The reddening distribution is almost flat for $E(J-K)<0.3$ ($E(B-V)<0.7$) mag with a slowly decreasing tail to higher values. The MWSC survey is dominated by older clusters. More than 45\% of the clusters have ages between 400~Myr and 2~Gyr. This can be attributed to the NIR basis of the survey since the older clusters with red giants are sufficiently prominent even at large distances from the Sun. On the other hand, the local maxima in the age distribution both for extremely old and extremely young clusters are a consequence of including globular clusters on the one hand and dedicated samples of embedded NIR clusters on the other.

\section{General description of the cluster sample}\label{sec:smpl}

The spatial distribution of clusters from our MWSC survey in the Galactic $XY-$plane is shown in Fig.~\ref{fig:xy}. The plot includes both open and globular clusters (note that the five most distant globular clusters do not appear in Fig.~\ref{fig:xy} since they are outside the plotted frame). The open clusters cover a wide range of galactocentric distances and reach the central region of the Galaxy as well as its fringe at about 20 kpc from the Galactic centre.

\begin{figure}
\begin{center}
 \includegraphics[width=\hsize]{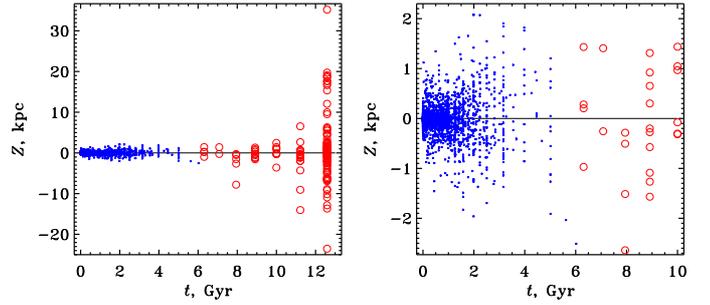}
 \caption{ Distribution of clusters along the Galactic $Z$-axis as a function of age. Open and globular clusters are marked with small blue  crosses and open red circles, 
respectively. The right panel shows  a zoomed-in segment of the left panel.}
   \label{fig:zage}
\end{center}
\end{figure}

Fig.~\ref{fig:kinl} illustrates the kinematic properties of open clusters selected in the MWSC survey within 4~kpc from the Sun. We show the distributions of proper motions and tangential and radial velocities versus Galactic longitude for different ranges of distances. As a reference, we also plot the systematic components of the velocities due to solar motion and Galactic rotation for heliocentric distances given at the top of each panel. The curves were computed with the corresponding parameters from COCD \citep[see][]{clupop} based on the Hipparcos data.   The top row of Fig.~\ref{fig:kinl} shows that the cluster proper motions coincide well with the predicted proper motion distribution. The systematic contribution of solar motion is relevant only for nearby clusters ($d_{XY} = d\cdot\cos b \ll$ 1.0~kpc). At larger distances the main systematic contributor is the Galactic rotation, and, on average, the observed proper motions of clusters follow this closely. This behaviour can be regarded as independent proof that cluster proper motions from the MWSC survey are in the inertial system represented by Hipparcos. The tangential velocities (the middle row of Fig.~\ref{fig:kinl}) and RVs (the bottom row of Fig.~\ref{fig:kinl}) also show an expected behaviour, which confirms that the distances and RVs of clusters in the MWSC survey are sufficiently reliable and can be used for studies of the Galactic disk kinematics. However, with increasing distances ($d_{XY} >$ 2.0~kpc) we found stronger deviations of the observed velocities from the simple rotation model, especially for proper motions and tangential velocities. This effect is probably also present in the RVs, but the number of data points is too low for a conclusive discussion. We cannot exclude the possibility that the accuracy of the cluster parameter determination may decrease at larger cluster distances, or that neglecting the radial metallicity gradient in the disk can bias the cluster distances. On the other hand, this effect may indicate that the adopted aproximation of the rotation law is valid for the solar neighbourhood, but is not sufficient to describe the Galactic rotation at larger distances from the Sun.

Completeness is one of the most important characterictics of a sample used in a statistical study. In Fig.~\ref{fig:sdprof} we show the surface density of open clusters in the MWSC survey as a function of their distance $d_{XY}$ in the Galactic plane. The upper panel shows young ($\log t < 7.9$, blue curve) and moderately young clusters ($\log t=7.9\dots8.3$, green curve), the bottom panel moderately old ($\log t=8.3\dots9.0$, brown curve) and old clusters ($\log t>9.0$, red curve). Additionally, we plot the total distribution of all open clusters in both panels (black curve). The total distribution is almost flat up to 1.8 kpc and steadily declines towards larger distances. This shows the increasing incompleteness of the MWSC survey at projected distances larger than 1.8~kpc. The average surface density of clusters with $d_{XY} <$ 1.8~kpc is 121 kpc$^{-2}$, which is close to the 114 kpc$^{-2}$ determined previously from the sample of COCD clusters within the smaller completenes area of 0.85 kpc \citep{clupop}.

The distributions of clusters of different ages follow the general trend, although with a few exceptions. A significant excess of the youngest clusters at about 400 pc from the Sun is related to the Orion star formation complex. For the youngest and oldest clusters the distributions decrease more slowly at $d_{XY} >$ 1.8~kpc because these clusters may contain absolutely bright stars, and therefore they can be observed in the NIR at larger distances than clusters of moderate ages. Another prominent feature appears in the distribution of the oldest (red curve in the bottom panel of Fig.~\ref{fig:sdprof}) clusters: the surface density of nearby clusters increases with increasing $d_{XY}$ up to distances of about 1.1 kpc from the Sun. This effect hints at an incompleteness of the input list at high Galactic latitudes. Indeed, one of the most important sources of input data for the MWSC survey was the cluster list by \citet{froeb07} obtained from a systematic search for clusters in the 2MASS catalogue at low Galactic latitudes $|b|<20^\circ$. Due to the strong concentration of younger clusters to the Galactic plane, this limitation has no serious consequences for their completeness in the MWSC survey. The oldest clusters show a larger scale height and a larger scattering in the $Z$-direction however (see below). Therefore, they could be incompletely represented in the MWSC survey, especially in the solar neighbourhood. Assuming that the real surface density of the oldest clusters is equal to the average density observed in the range $d_{XY}=1.1\dots2$ kpc, we estimate that $\approx 40$ old clusters are missing within  $d_{XY}\leqslant 1$ kpc. This effect must be taken into account in statistical studies based on the MWSC survey.

The cluster distribution from the MWSC survey along the Galactic $Z$-axis is shown in Fig.~\ref{fig:zage} as a function of cluster age. Here we consider both open and globular clusters with the parameters determined in this paper. The most striking feature of the distribution is the smooth transition from the oldest open clusters to the youngest globulars at $\approx 6$~Gyr. Within the age range $t\approx3\dots7$~Gyr, open and globular clusters show a similar vertical distribution. For older globular clusters, the vertical scattering increases steadily with age.   

\section{Summary and conclusions}\label{sec:conc}

Our MWSC project aimed at completing a full survey of star clusters in the wider neighbourhood of the Sun. We provided a catalogue containing the basic astrophysical data for all clusters in this survey, that is, exact positions of the cluster centre, proper motions, apparent radii, distances, reddenings, and ages. All these quantities were determined from data in the stellar all-sky surveys PPMXL and 2MASS. Because these latter surveys give homogeneous data sets all over the sky, and because a uniform pipeline has been applied to all objects, the astrophysical quantities derived in this paper have a uniform and homogeneous nature. A reliable membership determination is the basic issue; all the derived astrophysical parameters rely on this.

The starting point of the MWSC survey was an input list of targets compiled from the literature. The 3784 objects in the input list were called named clusters throughout the paper. From our analysis based on PPMXL and 2MASS data, we found that 3006 (79\,\%) of the named clusters are related to real objects. The vast majority of confirmed objects are open clusters, but due to the content of the input list, stellar associations and globular clusters were also found. In addition to the basic astrophysical parameters mentioned above, we also determined tidal parameters for the bulk of the clusters (98\%). From RV measurements of cluster members available in the literature, we estimated additional mean radial velocities for more than 30\% of the MWSC clusters.

For about 1500, or 50\,\% of the confirmed MWSC clusters, we presented basic astrophysical parameters for the first time. We also compared our results with literature data for subsets of clusters where possible. No severe systematic differences were found. Considering the size of the cluster sample and the uniform and homogeneous nature of the cluster parameters, the  MWSC survey is unprecedented.

Our sample of MWSC clusters covers a large section of the Galactic disk and reaches the very centre of the Milky Way as well as its outer regions. The area of data completeness now reaches the neighbouring spiral arms, which allows comparative studies of the cluster population in the inter- and intra-arm regions. Our sample of open clusters is almost complete up to a distance of about 1.8 kpc from the Sun, except for the subset of the oldest open clusters ($\log t \gtrsim 9$), where we found evidence of incompleteness within 1 kpc from the Sun. We attribute this effect to the incompleteness of our input list, that is, the data from the literature. This stimulated a search for clusters missing in the literature. This work is in progress at the moment of writing, but we will soon be able to publish a list that completes the current data.

\begin{acknowledgements}
This study was supported by DFG grant RO 528/10-1, and  by Sonderforschungsbereich SFB 881 "The Milky Way System" (subproject B5) of the German Research Foundation (DFG). We acknowledge the use of the Simbad database, the VizieR Catalogue Service and other services operated at the CDS, France, and the WEBDA facility, operated at the Institute for Astronomy of the University of Vienna. We thank the referee, Bruce A. Twarog, for his useful comments.
\end{acknowledgements}

\bibliographystyle{aa}
\bibliography{/home/tolya/bibdbs/clubib}

\begin{thebibliography}{26}
\expandafter\ifx\csname natexlab\endcsname\relax\def\natexlab#1{#1}\fi

\bibitem[{{Ahn} {et~al.}(2012){Ahn}, {Alexandroff}, {Allende Prieto},
  {Anderson}, {Anderton}, {Andrews}, {Aubourg}, {Bailey}, {Balbinot}, {Barnes},
  \& et~al. {et al.}}]{ahn12}
{Ahn}, C.~P., {Alexandroff}, R., {Allende Prieto}, C., {et~al.} 2012, \apjs,
  203, 21

\bibitem[{Bica {et~al.}(2003)Bica, Dutra, Soares, \& Barbuy}]{bicaea03}
Bica, E., Dutra, C., Soares, J., \& Barbuy, B. 2003, A\&A, 404, 223

\bibitem[{{Bica} {et~al.}(2003){Bica}, {Dutra}, \& {Barbuy}}]{bicadb03}
{Bica}, E., {Dutra}, C.~M., \& {Barbuy}, B. 2003, \aap, 397, 177

\bibitem[{{Bronfman} {et~al.}(1996){Bronfman}, {Nyman}, \& {May}}]{bron96}
{Bronfman}, L., {Nyman}, L.-A., \& {May}, J. 1996, \aaps, 115, 81

\bibitem[{{Bukowiecki} {et~al.}(2011){Bukowiecki}, {Maciejewski}, {Konorski},
  \& {Strobel}}]{bukea11}
{Bukowiecki}, {\L}., {Maciejewski}, G., {Konorski}, P., \& {Strobel}, A. 2011,
  \actaa, 61, 231

\bibitem[{{Conrad} {et~al.}(2013){Conrad}, {Scholz}, {Kharchenko}, {Piskunov},
  {Schilbach}, \& {R{\"o}ser}}]{conrad13}
{Conrad}, C., {Scholz}, R.-D., {Kharchenko}, N.~V., {et~al.} 2013, A\&A, 
submitted

\bibitem[{Dias {et~al.}(2002)Dias, Alessi, Moitinho, \& L{\'e}pine}]{daml02}
Dias, W.~S., Alessi, B.~S., Moitinho, A., \& L{\'e}pine, J. R.~D. 2002, A\&A,
  389, 871

\bibitem[{Dutra {et~al.}(2003)Dutra, Bica, Soares, \& Barbuy}]{dubisb03}
Dutra, C., Bica, E., Soares, J., \& Barbuy, B. 2003, A\&A, 400, 533

\bibitem[{{Froebrich} {et~al.}(2010){Froebrich}, {Schmeja}, {Samuel}, \&
  {Lucas}}]{froea10}
{Froebrich}, D., {Schmeja}, S., {Samuel}, D., \& {Lucas}, P.~W. 2010, \mnras,
  409, 1281

\bibitem[{Froebrich {et~al.}(2007)Froebrich, Scholz, \& Raftery}]{froeb07}
Froebrich, D., Scholz, A., \& Raftery, C.~L. 2007, MNRAS, 374, 399

\bibitem[{Harris(1996)}]{harris96}
Harris, W.~E. 1996, AJ, 112, 1487

\bibitem[{{Harris}(2010)}]{harcat10}
{Harris}, W.~E. 2010, ArXiv e-prints

\bibitem[{{Isobe} {et~al.}(1990){Isobe}, {Feigelson}, {Akritas}, \&
  {Babu}}]{isobe90}
{Isobe}, T., {Feigelson}, E.~D., {Akritas}, M.~G., \& {Babu}, G.~J. 1990, \apj,
  364, 104

\bibitem[{{Kharchenko} {et~al.}(2005{\natexlab{a}}){Kharchenko}, {Piskunov},
  {R{\"o}ser}, {Schilbach}, \& {Scholz}}]{clucat}
{Kharchenko}, N.~V., {Piskunov}, A.~E., {R{\"o}ser}, S., {Schilbach}, E., \&
  {Scholz}, R.-D. 2005{\natexlab{a}}, \aap, 438, 1163

\bibitem[{{Kharchenko} {et~al.}(2005{\natexlab{b}}){Kharchenko}, {Piskunov},
  {R{\"o}ser}, {Schilbach}, \& {Scholz}}]{newc109}
{Kharchenko}, N.~V., {Piskunov}, A.~E., {R{\"o}ser}, S., {Schilbach}, E., \&
  {Scholz}, R.-D. 2005{\natexlab{b}}, \aap, 440, 403

\bibitem[{{Kharchenko} {et~al.}(2012){Kharchenko}, {Piskunov}, {Schilbach},
  {R{\"o}ser}, \& {Scholz}}]{khea12}
{Kharchenko}, N.~V., {Piskunov}, A.~E., {Schilbach}, E., {R{\"o}ser}, S., \&
  {Scholz}, R.-D. 2012, \aap, 543, A156

\bibitem[{{Kharchenko} {et~al.}(2007{\natexlab{a}}){Kharchenko}, {Scholz},
  {Piskunov}, {Roeser}, \& {Schilbach}}]{newrvc2}
{Kharchenko}, N.~V., {Scholz}, R.-D., {Piskunov}, A.~E., {Roeser}, S., \&
  {Schilbach}, E. 2007{\natexlab{a}}, VizieR Online Data Catalog, 3254, 0

\bibitem[{{Kharchenko} {et~al.}(2007{\natexlab{b}}){Kharchenko}, {Scholz},
  {Piskunov}, {Roeser}, \& {Schilbach}}]{newrvcm}
{Kharchenko}, N.~V., {Scholz}, R.-D., {Piskunov}, A.~E., {Roeser}, S., \&
  {Schilbach}, E. 2007{\natexlab{b}}, VizieR Online Data Catalog, 10, 32801

\bibitem[{{King}(1962)}]{king62}
{King}, I. 1962, \aj, 67, 471

\bibitem[{Lada \& Lada(2003)}]{lala03}
Lada, C.~J. \& Lada, E.~A. 2003, ARA\&A, 41, 57

\bibitem[{{Melnik} \& {Dambis}(2009)}]{meldam09}
{Melnik}, A.~M. \& {Dambis}, A.~K. 2009, MNRAS, 400, 518

\bibitem[{{Monet} {et~al.}(2003){Monet}, {Levine}, {Canzian}, {Ables}, {Bird},
  {Dahn}, {Guetter}, {Harris}, {Henden}, {Leggett}, {Levison}, {Luginbuhl},
  {Martini}, {Monet}, {Munn}, {Pier}, {Rhodes}, {Riepe}, {Sell}, {Stone},
  {Vrba}, {Walker}, {Westerhout}, {Brucato}, {Reid}, {Schoening}, {Hartley},
  {Read}, \& {Tritton}}]{usnob1}
{Monet}, D.~G., {Levine}, S.~E., {Canzian}, B., {et~al.} 2003, \aj, 125, 984

\bibitem[{Piskunov {et~al.}(2006)Piskunov, Kharchenko, R{\"o}ser, Schilbach, \&
  Scholz}]{clupop}
Piskunov, A.~E., Kharchenko, N.~V., R{\"o}ser, S., Schilbach, E., \& Scholz,
  R.-D. 2006, A\&A, 445, 545

\bibitem[{{R{\"o}ser} {et~al.}(2010){R{\"o}ser}, {Demleitner}, \&
  {Schilbach}}]{ppmxl}
{R{\"o}ser}, S., {Demleitner}, M., \& {Schilbach}, E. 2010, \aj, 139, 2440

\bibitem[{{Siess} {et~al.}(2000){Siess}, {Dufour}, \& {Forestini}}]{siessea00}
{Siess}, L., {Dufour}, E., \& {Forestini}, M. 2000, \aap, 358, 593

\bibitem[{{Skrutskie} {et~al.}(2006){Skrutskie}, {Cutri}, {Stiening},
  {Weinberg}, {Schneider}, {Carpenter}, {Beichman}, {Capps}, {Chester},
  {Elias}, {Huchra}, {Liebert}, {Lonsdale}, {Monet}, {Price}, {Seitzer},
  {Jarrett}, {Kirkpatrick}, {Gizis}, {Howard}, {Evans}, {Fowler}, {Fullmer},
  {Hurt}, {Light}, {Kopan}, {Marsh}, {McCallon}, {Tam}, {Van Dyk}, \&
  {Wheelock}}]{cat2MASS}
{Skrutskie}, M.~F., {Cutri}, R.~M., {Stiening}, R., {et~al.} 2006, \aj, 131,
  1163

\end{thebibliography}

\end{document}